\documentclass{aastex631}

\graphicspath{{./}{figures/}}

\submitjournal{ApJL}

\usepackage{nameref}
\usepackage{longtable}

\shorttitle{Jupiter's Metastable Companions}
\shortauthors{Greenstreet et al.}

\begin{document}

\title[Transient Jovian Co-orbitals]{Jupiter's Metastable Companions}

\correspondingauthor{Sarah Greenstreet}
\email{sarahjg@uw.edu, sarah.greenstreet@noirlab.edu}

\author[0000-0002-4439-1539]{Sarah Greenstreet}
\affiliation{Department of Astronomy \& the DiRAC Institute, University of Washington, 3910 15th Ave NE, Seattle, WA 98195}
\affiliation{Rubin Observatory / NSF's NOIRLab, 
950 N. Cherry Ave, Tucson, AZ 85719}

\author[0000-0002-0283-2260]{Brett Gladman}
\affiliation{Department of Physics and Astronomy,  University of British Columbia, 6224 Agricultural Rd, Vancouver, B.C., V6T 1Z1}

\author[0000-0003-1996-9252]{Mario Juri\'c}
\affiliation{Department of Astronomy \& the DiRAC Institute, University of Washington, 3910 15th Ave NE, Seattle, WA 98195}

\begin{abstract}

Jovian co-orbitals share Jupiter's orbit and exhibit 1:1 mean motion resonance with the planet. 
This includes $>$10,000 so-called Trojan asteroids surrounding the leading (L4) and trailing (L5) Lagrange points, viewed as stable groups dating back to planet formation.
A small number of extremely transient horseshoe and quasi-satellite co-orbitals have been identified, which only briefly ($<$1,000 years) 
exhibit co-orbital motions.
Via a massive numerical study 
we identify for the first time some Trojans 
which are certainly only `metastable';
instead of being primordial, they are 
recent captures from heliocentric orbits into moderately long-lived (10~kyr-100~Myr) 
metastable states that will escape 
back to the scattering regime.  
We have also identified (1) the first two jovian horseshoe co-orbitals that 
exist for many resonant libration periods,
and (2) eight jovian quasi-satellites with metastable lifetimes of 4--130 kyr.
Our perspective on the Trojan population is thus now more complex as Jupiter joins the other giant planets in having known metastable co-orbitals which are in steady-state equilibrium with the planet-crossing Centaur and asteroid populations; the 27 identified here are in agreement with theoretical estimates.

\end{abstract}




\section{Introduction} \label{sec:intro}

The five famous Lagrange points of the 
circular restricted 
three-body problem are locations relative to the moving planet where objects have tiny relative accelerations.
In particular the `triangular' L4 and L5 Lagrange points are located 60 degrees ahead of and behind the planet along its orbit, and small bodies can oscillate
for long durations back and forth around these points. 
The L4/L5 stability was initially a theoretical discovery, which was followed by the first Trojan detections in 1906 \citep{Nicholson1961,Shoemakeretal1988},
but now include more than 10,000 cataloged members; 
these $>$10,000 Trojans are viewed as stable populations that date back to planet formation. 

Twenty-five years ago, \cite{Levisonetal1997} computed the
stability of the first 270 Jupiter Trojans on their nominal orbits, showing that some Trojans may leave in the next 0.3--4 billion years; that study assumed all Trojans were primordial and that any recent departures were due to a combination of
collisions and dynamical erosion, allowing some primordial
Trojans to leak away at the current epoch.
Here we will demonstrate the additional importance of recent temporary (metastable) captures into and out of co-orbital states on the shorter time scales of tens of kyr to Myr. 

Most planets are now known to host temporary
co-orbital companions (reviewed in \cite{Greenstreetetal2020} and \cite{Alexandersenetal2021}),
defined as objects undergoing oscillation (libration) of their
1:1 resonant argument 
for time scales much shorter than the age of the Solar System before escaping the resonance;
for direct orbits 
the resonant argument 
is simply the angle between the mean longitudes of the objects and planet. 
In addition to Trojans, co-orbital motion can be of 
the horseshoe type (when the small body passes through the direction $180^o$ away from the planet and motion encloses both the L4 and L5 points). Like Trojan motion, horseshoe orbits were predicted analytically, but are in most cases very unstable \citep{Rabe1961}. 
No long-term stable horseshoe sharing a planet's solar 
orbit 
has ever been observed. 
Lastly, in the frame co-rotating with Jupiter,
so-called `quasi-satellites' have orbits that maintain
large-distance motion encircling the planet
\citep{Wiegertetal2000}. 

Restricting our attention to the giant planets, Uranus
and Saturn do not have L4 and L5 points stable for 4 Gyr 
\citep{NesvornyDones2002}. 
Nevertheless a metastable uranian L4 Trojan \citep{Alexandersenetal2013} 
and a metastable saturnian horseshoe orbit \citep{Alexandersenetal2021} 
are known; `metastable' objects are here defined by undergoing many resonant argument librations before exiting the co-orbital state.
Neptune's L4 and L5 points have long-term stability,
but both stable and metastable Neptune Trojans are known 
\citep{HornerLykawka2012,Linetal2021}. 
Curiously, Jupiter has been the sole giant planet to have no known metastable co-orbitals, despite the expectation that the planet should host such a population \citep{Greenstreetetal2020}.

\begin{figure}[ht!]%
\centering
\includegraphics[width=0.8\textwidth]{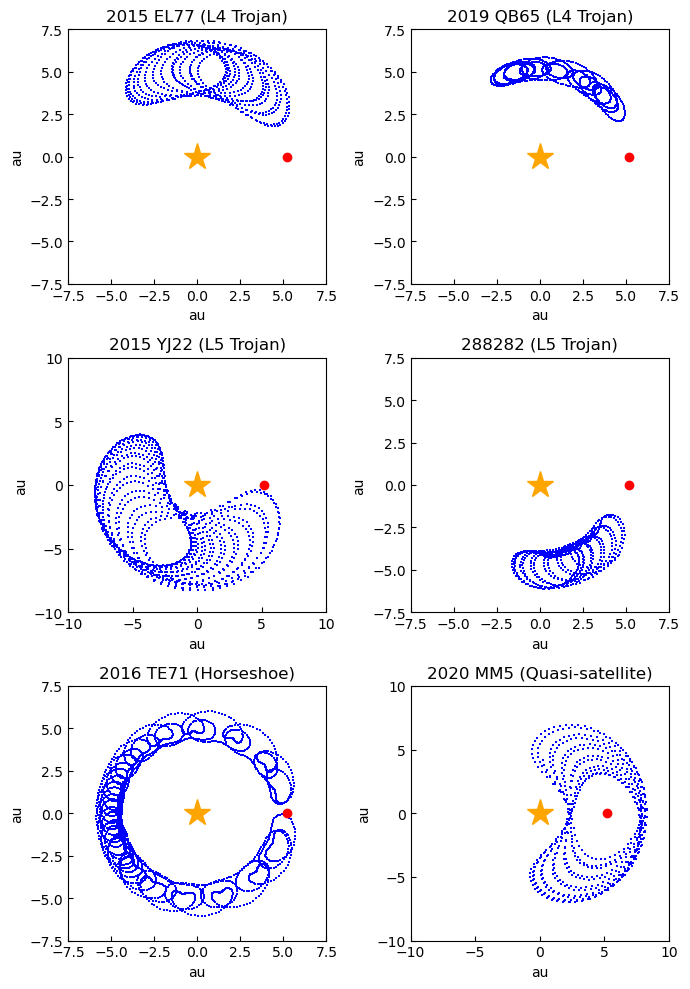}
\caption{Forward-integrated motion of 6 example metastable jovian co-orbitals identified in our analysis. Each object's motion is shown in the heliocentric reference frame co-rotating with Jupiter (red dot) for a single resonant libration period, which are roughly as follows: 2015 EL77 (L4): 165~yr, 2019 QB65 (L4): 175~yr, 2015 YJ22 (L5): 250~yr, 288282 (L5): 145~yr, 2016 TE71 (HS): 480~yr, 
 \& 2020 MM5 (QS): 145~yr. Note the different axis scales for each object.
}\label{fig_co-rotating}
\end{figure}

Planet-crossing small bodies can (rarely) find their way into
co-orbital states, and numerical simulations 
can estimate both the steady-state fraction relative to the 
current planet-crossing population and the expected distribution
of temporary-capture time scales (see Discussion). 
Because Jupiter is constantly being approached by objects
originating in the outer Solar System (Centaurs, that become Jupiter Family comets),
and given the estimated number of Jupiter-encountering Centaurs,
\cite{Greenstreetetal2020} calculated that 
the metastable capture fraction was high enough that 
metastable jovian co-orbitals should exist and trapping would generate all of Trojan, horseshoe, and quasi-satellite motions. Examples of all of these types will be illustrated in our
results below (see Figure~\ref{fig_co-rotating}). 

There has been a great deal of work studying the complex problem of co-orbital companions \citep{Christou1999,MoraisNamouni2013a,MoraisNamouni2013b,MoraisNamouni2019,Karlsson2004,BeaugeRoig2001,WajerKrolikowska2012,Wiegertetal2017,DiRuzzaetal2023}; these studies have either been done in the context of a simplified problem (one planet, sometimes on a circular orbit) or for time scales that are only slightly longer than the resonant libration period (of hundreds of years) or did not explore the range of behaviors and time scales possible due to the orbital uncertainties. 
Our work  pushes the sample size and the level of the model detail much further
by using full N-body simulations, by exploring time scales covering thousands of resonant libration periods, and by utilizing large numbers of `clones' drawn from the orbital uncertainty region for determining the  robustness of the resonant states; 
we also study the entire population of known objects with semimajor axes near that of Jupiter (nearly 12,000 objects). 
As a result, 
we have identified not only the first such metastable (2-13~kyr) jovian horseshoe orbits, but also 
the first known 
set of jovian Trojans
which are metastable on intermediate time scales of 0.01-30~Myr and must be recently captured into L4 or L5 motion, 
increasing the complexity of how we should view the Jupiter Trojan population. 


\begin{figure}[ht!]%
\centering
\includegraphics[width=1.0\textwidth]{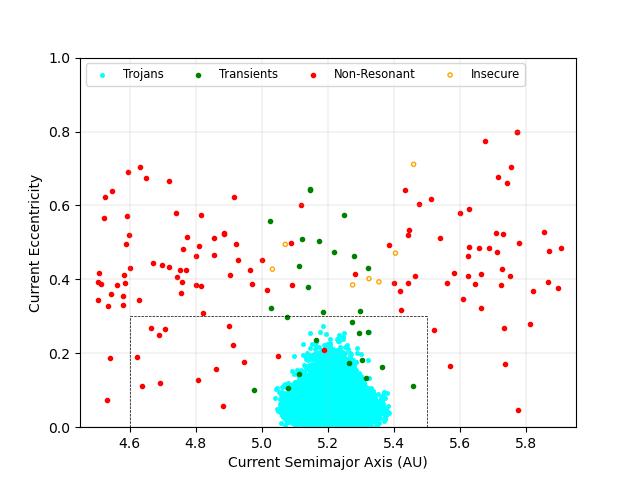}
\caption{Osculating semimajor axis vs eccentricity for the 11,581 objects with $a \simeq a_J = 5.20$~au that we classify with numerical integrations. 
The  11,423 ``Trojans" (cyan) are objects for which $\geq$95\% of their 1000 clones remain in 1:1 jovian resonance for 0.5~Myr. 
The 27 ``Transients" (green) have $\geq$95\% of their clones remain resonant for $\geq$1~kyr, but then leave the resonance (see Table~\ref{table:co-orb_classes}). 
The 124 ``Non-Resonant" (red) objects have $\geq$95\% of their 1000 clones ejected from the resonance in $<$1~kyr (see Table~\ref{table:non-res}).
The 7 ``Insecure" (orange) objects have 5\%-95\% of their 1000 clones remain in the resonance for $\geq$1~kyr before escaping (i.e., these objects would likely move to either ``non-resonant" or ``transient co-orbitals" upon further improvement of their orbital uncertainties; see Table~\ref{table:insecure}). 
The dashed rectangle shows the $a,e$ region that JPL Horizons and the Minor Planet Center (personal communication, Peter Vere\u{s}) 
currently define as the `jovian Trojan' parameter space; 
the 14 non-resonant objects 
in this box (listed in Table
~\ref{table:non-res}) 
are not Trojans, however,
given that $\geq$95\% of their 1000 numerically-integrated clones are ejected from the resonance in as little as tens or hundreds of years. 
The 27 metastable transients (green) have a larger range of semimajor axes, eccentricities, and inclinations
(see Figure 
\ref{fig_ai_ei} for the semimajor axis vs inclination and eccentricity vs inclination projections)
than the stable Trojans (cyan); objects can become temporarily bound to the resonance along its borders that stretch beyond the stable L4/L5 regions (cyan). 
\label{fig_ae}}
\end{figure}


\section{Methods}\label{sec_methods}

\subsection{Production of Sample Set and Dynamical Integrations}\label{subsec:sample_set_integrations}

To produce the sample set for this study, we queried the JPL Horizons Small Body Database Browser\footnote{\url{https://ssd.jpl.nasa.gov/tools/sbdb_query.html}} for objects fitting the following constraints: 
4.5$\leq$$a$$\leq$5.9~au (semimajor axis $a$ within twice Jupiter's Hill sphere radius), 
semimajor axis uncertainty, $a$-$sigma$, 
is defined, 
and the observational arclength, $data$-$arc$ $span$, 
is defined and is 
$>$30~days. We then manually removed all 
cometary provisional designations. In August 2022 this resulted in a sample set containing 11,581 known objects in the near-Jupiter region with arc-lengths of at least 30 days to ensure the orbital uncertainty was small enough to confidently be used to determine each object's orbital stability in the 1:1 co-orbital resonance with Jupiter.

We then used the Small Body Dynamics Tool (SBDynT)\footnote{\url{https://github.com/small-body-dynamics/SBDynT}} to query the JPL Horizons Small Body Database Browser to obtain the orbit and covariance matrix of each small body in our sample set, including for the best-fit orbit and 999 clones of each object within the orbital uncertainty region. This produced a set of 1000 ``clones" (best-fit orbit and 999 clones) for each of our 11,581 objects, totaling $\simeq11.6$ million state vectors. For more details on how our sample set is produced, see Section~\ref{subsec:app_sample_set} below).

To date, we have numerically integrated the 1000 clones for {\it all} of the 11,581 objects for 0.5~Myr into the future using the SWIFT-RMVS4 package \citep{Levisonetal1994}. In the 
Venus-Neptune 
planetary input files, we expanded Jupiter's radius by 1000x and turned on the ``lclose" exit condition in SWIFT to remove particles from the integrations when they come too close to Jupiter. In this study of 1:1 co-orbital resonant behavior, any particle that comes within 1000x Jupiter's radius ($\simeq$0.48~au, or about 2 Hill spheres) of the planet is unlikely to remain stable in the resonance and we terminate its integration. We use a base time step of 
3.7 days 
and an output interval of 1000 yrs. Particles are removed from the integrations when they get within 0.4 au or beyond 19.0 au from the Sun or too close to Jupiter as described above. We are currently continuing to integrate all $\simeq11.6$ million state vectors for longer time periods with the goal of eventually reaching 4~Gyr.


\section{Results}\label{sec_results}

We used the numerical integrations 
of the observationally-derived orbits and 999 clones within the orbital uncertainty region ($\simeq$11.6 million state vectors) to search for semimajor axis oscillation around Jupiter's value of 5.20~au as well as resonant argument libration for periods of time long enough ($>$1~kyr) to distinguish transient co-orbital capture or non-resonant behavior from primordial Trojan stability \citep{Greenstreetetal2020,Alexandersenetal2013,Alexandersenetal2021}. 
This calculation required approximately 20 CPU years on a Beowulf cluster at the University of British Columbia. 
Details of the methods used for co-orbital detection, resonant island classification, and determination of resonant-sticking time scales can be found in Sections~\ref{subsec:app_res_island_class} and \ref{subsec:app_res_sticking_times}.

We securely identify the transient co-orbitals and non-resonant objects in the sample of 11,581 objects 
in the `near-Jupiter population' 
(i.e., semimajor axes $a=$4.5--5.9~au, within $\simeq2$ jovian Hill sphere radii of Jupiter's $a_J$). 
We classify objects as belonging to one of the following dynamical classes based on their fraction of resonant clones and resonant time scales: ``Trojans", ``Transients", ``Non-Resonant", or ``Insecure" (see Figure~\ref{fig_ae} caption, Sections~\ref{subsec:app_res_island_class} and \ref{subsec:app_res_sticking_times}, and Table~\ref{table:num_in_class} for details). Figure~\ref{fig_ae} shows the semimajor axis vs eccentricity distribution of the sample of near-Jupiter objects along with our classifications. 
The ``Trojans" (objects for which $\geq$95\% of the 1000 clones remain in the 1:1 jovian resonance for 0.5~Myr) are deemed long-term stable and have not been integrated beyond this time scale; in the future we will extend these integrations to 4~Gyr to study the stability of these objects on Solar System time scales. All other objects (``transient", ``non-resonant", and ``insecure") have been integrated for time scales long enough that all 1000 clones have left the resonance; 
these integration time scales range from a few hundred years for the non-resonant objects to up to $\sim$2~Gyr for the transient co-orbitals. 
We note that some ``transient" or ``insecure" objects can become trapped in the 1:1 resonance multiple times during the integrations. 
We base our classifications on the start of the integrations (i.e., the current time) and do not discuss (rare) multiple resonant traps in this paper. 

Among the near-Jupiter sample, we have identified 27 objects (Table~\ref{table:co-orb_classes}), 
which we are confident are \textit{not} primordial objects. 
Instead, they are almost certainly recently captured as Jupiter co-orbitals that remain metastable for time scales of $10^3-10^8$~years. 
While each of these 27 objects share a commonality with the primordial Trojans by their presence in the 1:1 jovian mean-motion resonance, they are unique in their much shorter resonant stability time scales that can only mean that they are recent captures into the co-orbital population and are thus required to be placed in a category of jovian co-orbitals separate from the primordial Trojans.

\begin{deluxetable}{cccccc}[ht]
\tablecaption{Classifications of metastable jovian co-orbitals. Objects in {\bf bold} are shown in Figures~\ref{fig_co-rotating} and/or \ref{fig_duration}. Table~\ref{table:res_escape_times} provides the resonance escape time scales for these 27 objects.}\label{table:co-orb_classes}%
\tablehead{
\colhead{Classification} &  \colhead{Members} &  \colhead{} &   \colhead{} &  \colhead{} &  \colhead{} \\
}
\startdata
L4 Trojan & 163240 & 2010 AQ134 & 2010 VT278 & 2014 EJ166 & {\bf 2015 EL77} \\
 & 2015 HF178\tablenotemark{1} & 2015 HX159 & 2017 PC52 & {\bf 2019 QB65} & 2020 RL50\\
 & 2020 RO89  & 2020 SN84 & & & \\
L5 Trojan & {\bf 288282} & 613709 & {\bf 2015 YJ22}\tablenotemark{1} & 2018 BE7 & \\
Horseshoe & 2015 OL106 & {\bf 2016 TE71} & & & \\
Quasi-satellite & 241944\tablenotemark{2} & 363135\tablenotemark{2} & 526889\tablenotemark{2} & 2003 WG133 & 2004 AE9\tablenotemark{2} \\
 & 2017 SN215 & 2018 UH25 & {\bf 2020 MM5}\tablenotemark{2} & & \\
Retrograde & {\bf 514107} & & & & \\
\enddata
\tablenotetext{1}{Over the next 600 years, \cite{DiRuzzaetal2023} classified 2015 HF178 as a Trojan (as we do) and 2015 YJ22 as a horseshoe (which we classify as a L5 Trojan), but we follow their orbital evolutions much longer and show that these objects leave the resonance and are not primordial.}
\tablenotetext{2}{\cite{WajerKrolikowska2012} and/or \cite{DiRuzzaetal2023} classified 241944, 363135, 526889, 2004 AE9, and 2020 MM5 as current quasi-satellites; \cite{WajerKrolikowska2012} integrated the nominal orbits of 241944 and 526889 for $\sim$10~kyr and identified their transient nature in the resonance and classified 353135 and 2004 AE9 as long-lasting quasi-satellites remaining stable for $>$10~kyr. By studying longer time scales, we find median resonant lifetimes for the 1000 clones of each of these 5 objects to be 20-130~kyr.}
\end{deluxetable}

\begin{figure}[ht]%
\centering
\includegraphics[width=1.0\textwidth]{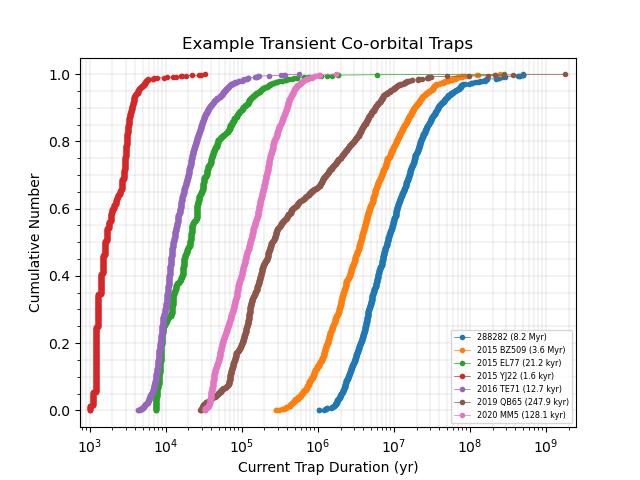}
\caption{Cumulative distribution for the resonance escape times for the 1000 clones of 7 selected transient jovian co-orbitals. The number in () after each designation is the median resonant 
time scale for each object's current trap in the 1:1 resonance. For the full list of resonant sticking time scales for each of our 27 identified metastable jovian co-orbitals, see Table~\ref{table:res_escape_times}.
}\label{fig_duration}
\end{figure}

First, we identify 12 L4 and 4 L5 Trojans (four of which are shown in Figures \ref{fig_co-rotating} and \ref{fig_duration}) that 
are surely unstable on time scales much shorter than 
ever previously discussed (only $\sim$Gyr time scales are 
discussed in \cite{Levisonetal1997}).  
The median time scales over which these metastable Trojans escape the resonance range from 1~kyr-23~Myr (Table~\ref{table:res_escape_times}), however, their observational uncertainties result in instability time scales that vary by an order of magnitude or more, as evidenced by the range in escape times of each object's 1000 clones (see Figure~\ref{fig_duration}). 
This rapid departure means these 16 L4/L5 metastable Trojans cannot be members of the primordial population which are departing today, but must be 
recent metastable captures.

To date, no transient horseshoe co-orbitals of Jupiter have been identified to librate in the resonance on time scales of more than a couple hundred years 
(long enough for the object to experience several libration periods), despite the expectation that they should  exist among  
the metastable jovian co-orbital population \citep{Greenstreetetal2020}. 
We have here identified the first two known metastable horseshoes of 
Jupiter: 2015 OL106 and 2016 TE71.  
The latter provides the first known example of a real object that remains in horseshoe motion with Jupiter for dozens of libration periods and resembles historical predictions of jovian horseshoe behavior \citep{Rabe1961}.  2016 TE71 is shown in Figures~\ref{fig_co-rotating} and \ref{fig_duration}. 

Altogether, our metastable identifications include 12 L4 Trojans, 4 L5 Trojans, 2 horseshoes, 8 quasi-satellites, and the retrograde jovian co-orbital (514107) Ka`epaoka`$\bar{\mathrm{a}}$wela 2015 BZ509. 
We note that a handful of these objects have been previously classified \citep{Karlsson2004,WajerKrolikowska2012,DiRuzzaetal2023} based on their dynamical behavior over the next few hundred to couple thousand years (see captions for Tables~\ref{table:co-orb_classes}, 
\ref{table:insecure}, 
and \ref{table:non-res}); 
differences between previous shorter time scale classifications and the longer metastable time scale classifications presented here are discussed 
below. 
Figure~\ref{fig_co-rotating} shows the forward-integrated motion 
for a single libration period for six metastable object examples. 

A unique aspect of our work is 
to determine the time scales over which these objects (and the clones representing their orbital uncertainty) will eventually escape the resonance. 
To determine their metastable time scales, 
we extended  each object's 
integrations until all 1000 clones were removed (most often for getting too close to Jupiter). 
The cumulative distributions for the resonance escape times for the 1000 clones of 
each of the six objects shown in Figure~\ref{fig_co-rotating} are given in  
Figure~\ref{fig_duration}. 
This figure also presents our measurement of the instability time scale 
for the retrograde co-orbital (514107) with median value of 3.6~Myr; 
\cite{Wiegertetal2017} estimated that the object remained in the near-Jupiter region for a lower limit of at least 1~Myr, while \cite{NamouniMorais2020} estimated a median lifetime of 6.5~Myr for the object to escape the Solar System or collide with the Sun. 
The examples shown in Figure~\ref{fig_duration} 
depict the range in metastable resonant sticking time scales ($10^3-10^8$ years or longer) we have identified so far. 
Table~\ref{table:res_escape_times} contains the full list of resonant sticking time scales for each of our 27 identified metastable jovian co-orbitals. For more details on the determination of the resonance escape time scales, see Section~\ref{subsec:app_res_sticking_times}.


\section{Discussion}\label{sec_discussion}

After the identification  of asteroid (514107) as a retrograde 
jovian co-orbital  \citep{Wiegertetal2017},
these are the first securely-identified  metastable
jovian co-orbitals for which the resonant sticking time scales have been established. 
While other groups \citep{BeaugeRoig2001,Karlsson2004,WajerKrolikowska2012,DiRuzzaetal2023} have identified resonant behaviors of some of these objects, those analyses do not extend beyond the next $\sim$1-10~kyr nor do they utilize large numbers of clones drawn from the orbital uncertainty region for determining the certainty of a resonant classification. 
We confirm the current non-resonant and quasi-satellite classifications for a handful of objects (see Tables~\ref{table:co-orb_classes} \& 
\ref{table:non-res}). However, our analysis is largely unique in identifying the transient nature of these objects by determining the time scales over which they (and the clones representing their orbital uncertainty) will eventually escape the resonance. 

We find a number of resonant classifications that differ from previous studies
\citep{BeaugeRoig2001,Karlsson2004,WajerKrolikowska2012,DiRuzzaetal2023}; these objects are noted in Tables~\ref{table:co-orb_classes}, \ref{table:insecure}, \& \ref{table:non-res}. 
Note that \cite{DiRuzzaetal2023} include many objects in their analysis having 
arc lengths of $\leq$5~days, which we 
omit
given our requirement that objects have arc lengths $>$30~days to ensure their orbital 
uncertainty regions are determined by the observations rather than dominated by orbit fitting assumptions. 
We additionally require resonant objects 
to librate in the 1:1 for at least 1~kyr in order to experience several libration periods before possible departure from the resonance (in the case of the transient captures). 
This is responsible for the classification differences for the objects that we classify as non-resonant that other studies find are resonant during the $<$1~kyr time scales they use (e.g., \cite{DiRuzzaetal2023} provide classifications based on 600~yr integrations). In addition, we integrate each object's 1000 clones until all the clones have been removed from the integrations, which allows us to securely classify each co-orbital as transient in nature and determine the time scales over which they are stable in the resonance. This differs from the majority of the previous studies, which can only determine if an object is currently resonant but not how long it will remain resonant 
nor the fact that observational uncertainties can result in instability time scales that vary by an order of magnitude or more (see Figure~\ref{fig_duration}). 

We expect the number of transient co-orbitals and primordial Trojans among the 11,581 object
sample to shift as we continue to integrate 
the 1000 clones for time periods longer than 0.5~Myr. 
Very long-lived resonant objects unstable  
in $\lesssim$1~Gyr (i.e., long-lived temporary captures) will become evident in longer integrations, shifting some objects from ``Trojan" to ``transient co-orbital" classification. 
This will then meld into the 
few long-known Jupiter Trojans 
unstable on Gyr time scales, which was suggested 
\citep{Levisonetal1997} to be due to a combination of
long-term dynamical erosion and collisions.

Our  perspective is  thus now  more complex.
The Jupiter co-orbital population consists of a mix of objects with different resonant time scales that 
we very loosely divide into the following categories: extremely transient ($\lesssim$1~kyr), metastable (10~kyr-100~Myr), primordial Trojan erosion ($\sim$Gyr), and stable Trojans (longer than 5~Gyr Solar System time scales). 
Cases of extremely transient objects, which only last one (or a few) resonant libration periods, have been studied
\citep[for example,][]{BeaugeRoig2001,Karlsson2004,WajerKrolikowska2012,DiRuzzaetal2023}. 
Here we have shown 
for the first time 
that Jupiter joins the other giant planets by having 
recently-trapped co-orbitals that last for an enormous range of metastable time scales 
(10 kyr - 100 Myr) 
consistent with the 
transient co-orbital populations of all the giant planets. 
At the very longest time scales, only Jupiter and Neptune harbor both
stable Trojan swarms and Trojans whose current stability time scales 
are of order Gyr.
These latter objects can be a combination of the longest-lived
traps of Centaurs and the slowly eroding edges of the original
primordial population.
The metastable objects we identify in this paper, however, must be recently captured into the co-orbital state out of the planet-crossing Centaur population, with a possible (probably small) contribution from escaping main-belt asteroids \citep{Greenstreetetal2020}. 

The metastable co-orbitals identified here thus represent the discovery
of the first (curiously-missing) jovian members
of the expected transient co-orbital population accompanying each giant planet \citep{Alexandersenetal2013,Greenstreetetal2020,Alexandersenetal2021}. 
Numerical simulations of the Centaur and escaped asteroid populations, both of which can become temporarily-trapped into 1:1 jovian resonance, allowed \cite{Greenstreetetal2020} to compute the steady-state fractions present in the jovian co-orbital population at any given time. 
Given the number of absolute magnitude $H<18$ (sizes of order 1~km) near-Earth objects (NEOs) and Centaurs \citep{Lawleretal2018}, \cite{Greenstreetetal2020} estimated 
that there should be $\sim$1-100 metastable jovian co-orbitals 
that remain resonant on time scales of $\lesssim$10~Myr. 
Here we identify 27 metastable jovian co-orbitals, all of which have $H<18$, that remain stable for time scales of $10^3-10^8$~years, in agreement with the theoretical estimate. 

More metastable jovian co-orbitals will certainly be telescopically detected; given the rarity of capture into co-orbital resonance these additional co-orbitals are likely to be small, which is partly the reason more have not been identified to date by current surveys. 
The upcoming Vera C.~Rubin Observatory Legacy Survey of Space and Time (LSST), with its large aperture and magnitude depth, should increase the number of Jupiter Trojan detections by $\sim$25x \citep{LSST2017}. 
These fainter detections will also provide more objects currently in metastable traps  with Jupiter; their identification as metasable, however, will require more than simple osculating element cuts in semimajor axis and eccentricity near Jupiter's values, 
as our massive numerical study has demonstrated (see Figure~\ref{fig_ae}). 

The Lucy spacecraft mission will visit five Trojans  during 2027 - 2033 \citep{Levisonetal2021}.  We have carefully integrated these Trojans for 50 Myr to study their stability in the 1:1 jovian resonance. We find that all 1000 clones for each of these five mission targets remain stable in the resonance over this time scale and thus are almost  certainly primordial objects. 

A preliminary examination of the (sparse) color data from
the Sloan Digital Sky Survey \citep{SergeyevCarry2021}
for the faint metastable co-orbitals identified here
shows that, relative to most known Trojans 
\citep{Szaboetal2007} 
and the Lucy flyby targets, the objects 
2016 TE71 (metastable horseshoe),
(288282) 2004 AH4 (metastable L5 Trojan),
and (163240) 2002 EM157 (metastable L4 Trojan) 
have evidence for redder photometric $g-r$ and/or
$g-i$ optical colors than typical Trojans.
This would
be expected if they are recently trapped Centaurs. 


\begin{acknowledgments}
This work is supported by NASA Solar System Workings grant 80NSSC22K0978. The integrations used for this work's massive numerical study were run on a Beowulf cluster at the University of British Columbia. This work utilized the Small Bodies Dynamics Tool (SBDynT) (\url{https://github.com/small-body-dynamics/SBDynT}) and the Keplerian-to-cartesian element conversion tool from THOR (Tracklet-less Heliocentric Orbit Recovery; \url{https://github.com/moeyensj/thor}). We wish to thank Kat Volk for allowing us access to and helping us to use the SBDynT for this work, Mike Alexandersen for sharing his co-rotating reference frame plotting script that was adapted for producing the relevant plots in this paper, and Pedro Bernardinelli for helpful discussions that improved the quality of the paper. 
BG acknowledges funding support from the Canadian Space Agency and NSERC.
\end{acknowledgments}


\newpage

\appendix

\section{Details on Dynamical Integrations}

\subsection{Sample Set Production}\label{subsec:app_sample_set}

After querying the JPL Horizons Small Body Database Browser\footnote{\url{https://ssd.jpl.nasa.gov/tools/sbdb_query.html}} to obtain the 11,581 near-Jupiter objects in our sample set and then using the Small Body Dynamics Tool\footnote{\url{https://github.com/small-body-dynamics/SBDynT}} to obtain the best-fit orbit and 999 clones within the orbital uncertainty region of each object, the SBDynT then queries Horizons for the value of $GM$ associated with each orbit and converts the best-fit and clone keplerian orbits to heliocentric cartesian positions and velocities. This produced a set of 1000 ``clones" (best-fit orbit and 999 clones) for each of our 11,581 objects, totaling $\simeq11.6$ million state vectors. Each object's set of 1000 clones were for a non-user-determined epoch associated with the Horizons-generated covariance matrix for a given object. In order to generate a set of planetary positions and velocities to be used for numerically integrating each object's clone set, we then used the SBDynT to obtain the heliocentric position and velocity of Venus--Neptune at the corresponding epoch from JPL Horizons via its web API request. 

\subsection{Co-orbital Detection and Resonant Island Classification}\label{subsec:app_res_island_class}

In order to distinguish transient co-orbital capture (or  even entirely non-resonant behavior) from primordial Trojan stability, we used a 1~kyr running window to search for semimajor axis oscillation around Jupiter's $a_J$=5.20~au, as well as resonant argument libration. 
Similar to the requirements used in \cite{Greenstreetetal2020} and \cite{Alexandersenetal2013,Alexandersenetal2021}, a particle's average semimajor axis must remain within 0.2~au of $a_J$ and no individual semimajor axis value may differ by no more than 0.65~au from $a_J$ within the 1~kyr window, if the object is to be considered resonant during that 1~kyr time period. 
Objects for which $<5$\% of the 1000 clones are resonant for at least a single 1~kyr time period, are classified as ``non-resonant". These include the 14 MPC/JPL-classified ``Trojans" we find for which $\geq95$\% of the 1000 clones leave the near-Jupiter region in as little as tens or hundreds of years, which are clearly \textit{not} long-term stable resonant L4/L5 primordial Trojans. We also identify another 110 asteroids and Centaurs that are currently not in co-orbital resonance with Jupiter (see Tables~\ref{table:num_in_class} and \ref{table:non-res}).

Objects for which $\geq95$\% of the 1000 clones are resonant for the duration of the 0.5~Myr integration, are classified as long-term stable ``Trojans". We identify 7482 L4 Trojans and 3941 L5 Trojans (11,423 Trojans in total) among our 11,581 object sample set. Objects that have $\geq95$\% of their 1000 clones resonate for at least 1~kyr but then leave the resonance are classified as ``transient co-orbitals". To date, we have identified 27 metastable jovian co-orbitals (see Table~\ref{table:res_escape_times}). We further classify the metastable co-orbitals by their resonant argument ($\phi_{11}=\lambda-\lambda_J$, where $\lambda$ is the object's mean longitude and $\lambda_J$ is that of Jupiter) behavior as L4 or L5 Trojans ($\phi_{11}$ remains in the leading or trailing hemisphere, respectively, during a 1~kyr running period), Horseshoes ($\phi_{11}$ crosses the direction $180^o$ away from Jupiter at any time during a window interval), or Quasi-satellites (all other objects since these must cross between the leading and trailing hemispheres at $\phi_{11}=0^o$ instead of $180^o$) \citep{Greenstreetetal2020,Alexandersenetal2013,Alexandersenetal2021}. Among our 27 identified metastable co-orbitals, we find 12 L4 Trojans, 4 L5 Trojans, 2 horseshoes, 8 quasi-satellites, and the retrograde co-orbital (514107) Ka`epaoka`$\bar{\mathrm{a}}$wela 2015 BZ509 (see Table~\ref{table:num_in_class}). 

A minor shortcoming of using a running window for diagnosing co-orbital behavior is that the end of each resonant object's current trap may not be well identified due to the end of the window not falling entirely within the trap. Additionally, the resonant island classification algorithm described above can produce erroneous classifications, in particular for co-orbitals with large-amplitude librations that encompass Lagrange points not typically associated with their resonant state (e.g., large-amplitude Trojans whose librations cross either the leading or trailing hemisphere at $\phi_{11}=180^o$ or $0^o$) or those that (rarely) transition to libration around another resonant island (e.g., a Trojan transitioning to a horseshoe). However, our algorithms are adjusted to account for these minor shortcomings \citep{Alexandersenetal2013,Greenstreetetal2020,Alexandersenetal2021}, and these errors affect $<$10\% of cases upon manual inspection of dozens of objects and their clones, including careful examination of the resonant behavior of the 27 metastable jovian co-orbitals we have identified among our 11,581-object sample set. 

Lastly, objects for which $5-95$\% of their 1000 clones remain resonant for at least a single 1~kyr running window period are classified as ``insecure". Upon further improvement of their orbital uncertainties, these objects would likely move from an ``insecure" classification to either ``non-resonant" or ``transient co-orbitals". We identify 7 ``insecurely" resonant objects among our sample set of 11,581 near-Jupiter objects; 
each of these 7 objects are classified by the MPC/JPL as asteroids (see Tables~\ref{table:num_in_class} \& \ref{table:insecure}). 

\subsection{Determination of Resonant Sticking Time Scales}\label{subsec:app_res_sticking_times}

Any object in our 11,581-object sample set found to have at least one clone that does not remain resonant for the duration of the 0.5~Myr integrations has had their integrations extended (or are being extended) to the point where all 1000 clones have been removed from the integrations by entering one of the exit criteria described above. 
The 27 metastable co-orbitals, 124 non-resonant objects, and 7 ``insecure" objects all fall into this category; none of 11,423 L4/L5 Trojans in our sample set had a single clone leave the 1:1 jovian resonance in the 0.5~Myr integrations. While the non-resonant objects had $<$5\% of their 1000 clones librate in the resonance for $\geq$1~kyr, some non-resonant objects did have some number of clones that remained resonant for at least that length of time. The number of transient clones for each non-resonant object and the maximum trap durations of those transient clones are listed in Table~\ref{table:non-res}, along with the same information for the ``insecure" objects listed in Table~\ref{table:insecure}.

These longer integrations have been run for tens to hundreds of Myr, or in one case $\sim$2~Gyr. The extension of these numerical integrations until the final clone has exited allows us, for the first time, to determine their resonant trapping time scales (see Tables~\ref{table:res_escape_times}, \ref{table:insecure}, \& \ref{table:non-res}, and Figure~\ref{fig_duration}). A resonant ``trap" includes the duration of consecutive 1~kyr running windows for which a state vector satisfies our resonant criteria described above. Some ``transient" or ``insecure" objects can become trapped multiple times during the integrations. We base our classifications on the start of the integrations (i.e., the current time) and do not discuss (rare) multiple resonant traps in this paper.

\begin{figure}[ht!]%
\centering
\includegraphics[width=0.85\textwidth]{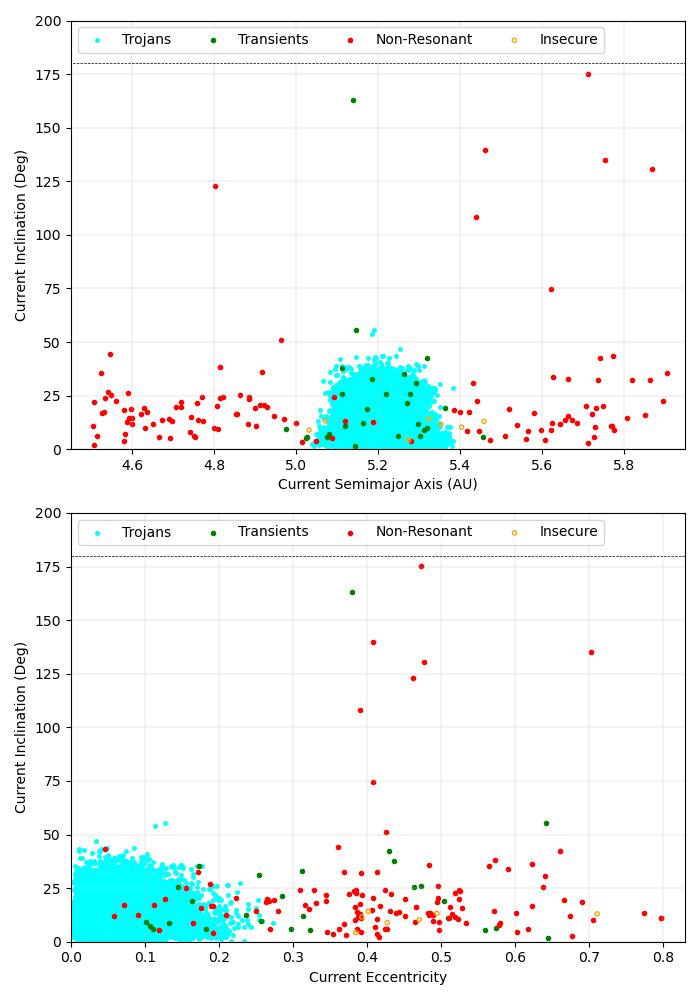}
\caption{Heliocentric semimajor axis vs osculating J2000 inclination (top) and ecccentricity vs inclination (bottom) for the 11,581 objects with $a \simeq a_J$ and their 1:1 jovian resonant classifications. (Similar to Figure~\ref{fig_ae}.)
}\label{fig_ai_ei}
\end{figure}

\begin{deluxetable}{cccccc}
\tablecaption{Resonant island configuration and resonance escape time scales for our 27 identified metastable ($10^3-10^8$~yr) jovian co-orbitals (i.e., objects we classify as ``transient" that have $\geq$95\% of their 1000 clones librate in the 1:1 jovian resonance for at least 1~kyr and are then ejected from the resonance (green points in Figures~\ref{fig_ae} \& \ref{fig_ai_ei}). Current resonant island configurations are classified as L4 Trojan, L5 Trojan, Horseshoe (HS), Quasi-satellite (QS), or Retrograde (R) motion. Min, median, and max trap durations refer to the amount of time the 1000 clones for each of these objects remain trapped in the 1:1 jovian resonance before being ejected. Objects in {\bf bold} are shown in Figures~\ref{fig_co-rotating} and/or \ref{fig_duration}.}\label{table:res_escape_times}
\tablehead{
\colhead{Designation} &  \colhead{JPL Class} &  \colhead{Island} &   \colhead{Min Trap (kyr)} &  \colhead{Median Trap (kyr)} &  \colhead{Max Trap (Myr)} \\
}
\startdata
(163240) 2002 EM157 & Trojan & L4 & 1928 & 22915 & $>$200\tablenotemark{1}   \\
(241944) 2002 CU147 & Asteroid & QS & 11.5  & 79   & 4.5   \\
{\bf (288282) 2004 AH4} & Trojan   & L5 & 1031  & 8192 & $>$500\tablenotemark{1}   \\
(363135) 2001 QQ199 & Asteroid & QS & 8.4   & 59   & 3.7   \\
{\bf (514107) 2015 BZ509} & Asteroid & R      & 281   & 3606 & 247   \\
(526889) 2007 GH6 & Asteroid & QS & 9.8   & 44   & 1.6   \\
(613709) 2007 RK185 & Asteroid & L5 & 1.2   & 1.2  & 0.006 \\
2003 WG166 & Asteroid & QS & 6.3   & 45   & 3.3   \\
2004 AE9   & Asteroid & QS & 14.5  & 19   & 0.8   \\
2010 AQ134 & Asteroid & L4 & 12.5  & 102  & 3.7   \\
2010 VT128 & Asteroid & L4 & 48.9  & 136  & 1.9   \\
2014 EJ166 & Trojan   & L4 & 1.2  & 7   & 1.8   \\
{\bf 2015 EL77}  & Trojan   & L4 & 7.4   & 21   & 285   \\
2015 HF178 & Asteroid & L4 & 5.5   & 9    & 1.1   \\
2015 HX159 & Trojan   & L4 & 1080  & 8704 & $>$200\tablenotemark{1} \\
2015 OL106 & Trojan   & HS & 1.7   & 2    & 0.5   \\
{\bf 2015 YJ22}  & Asteroid & L5 & 1.0  & 2   & 0.03   \\
{\bf 2016 TE71}  & Trojan   & HS       & 4.4   & 13   & 0.6   \\
2017 PC52  & Trojan   & L4 & 1.8   & 2    & 0.2   \\
2017 SN215 & Asteroid & QS & 1.9   & 4   & 0.5    \\
2018 BE7   & Trojan   & L5 & 5.9   & 25  & 2.2    \\
2018 UH25  & Asteroid & QS & 3.6   & 27  & 2.2    \\
{\bf 2019 QB65}  & Trojan   & L4 & 28.5  & 248 & 1818    \\
{\bf 2020 MM5}   & Asteroid & QS & 33.0  & 128 & 1.8    \\
2020 RL50  & Trojan   & L4 & 104.8 & 995  & 634  \\
2020 RO89  & Trojan   & L4 & 25.9  & 116 & 134.2  \\
2020 SN84  & Trojan   & L4 & 1.0  & 5   & 0.6   \\
\enddata
\tablenotetext{1}{(163240) 2002 EM157, (288282) 2004 AH4, and 2015 HX159 have 46, 4, 41 clones, respectively, still resonant at the end of 200~Myr, 500~Myr, and 200~Myr integrations, respectively.}
\end{deluxetable}

\begin{deluxetable}{cc}
\tablecaption{Number of objects in our 11,581 near-Jupiter object sample given each resonant classification.}\label{table:num_in_class}%
\tablehead{
\colhead{Resonant Configuration} &  \colhead{Number of Objects} \\
}
\startdata
\textbf{Transients}    & \textbf{27}     \\
L4 Trojans       & 12 \\
L5 Trojans       & 4 \\
Horseshoes       & 2    \\
Quasi-satellites & 8    \\
Retrograde       & 1    \\
\hline
\textbf{Non-Resonant}  & \textbf{124}    \\
Trojans       & 14   \\
Asteroids/Centaurs       & 110    \\
\hline
\textbf{Insecure}      & \textbf{7}     \\
Asteroids/Centaurs       & 7    \\
\hline
\textbf{Trojans}      & \textbf{11423}     \\
L4 Trojans    & 7482   \\
L5 Trojans    & 3941   \\
\hline
\textbf{Total}         & \hspace{-1.5mm}\textbf{11581}  \\
\enddata
\end{deluxetable}

\begin{deluxetable}{cccc}
\tablecaption{Objects we classify as ``insecure", i.e., 5-95\% of their 1000 clones remain resonant for $\geq1$~kyr and then leave the resonance (these objects would likely move to either ``non-resonant" or ``transient co-orbitals" upon further improvement of their orbital uncertainties; orange 
points in Figures~\ref{fig_ae} 
\& \ref{fig_ai_ei}). The number of transient clones that remain trapped in the 1:1 jovian resonance for $\geq$1~kyr and then leave the resonance are provided along with the maximum resonant trap durations.}\label{table:insecure}%
\tablehead{
\colhead{Designation} &  \colhead{JPL Class} & \colhead{Transient Clones} &   \colhead{Max Trap Duration (Myr)} \\
}
\startdata
(32511) 2001 NX17\tablenotemark{1} & Asteroid & 534 & 1.1 \\
2009 BM124 & Asteroid & 115 & 0.002 \\
2012 BL173 & Asteroid & 804 & 0.24 \\
2013 GE26 & Asteroid & 633 &  0.29 \\
2013 VB17 & Asteroid & 910 & 0.42 \\
2014 DB64 & Asteroid & 77 & 0.28 \\
2016 VS10 & Asteroid & 76 & 0.32 \\
\enddata
\tablenotetext{1}{\cite{Karlsson2004} classified the nominal orbit of (32511) 2001 NX17 as non-resonant.}
\end{deluxetable}

\startlongtable
\begin{deluxetable}{cccc}
\tablecaption{Objects we classify as ``non-resonant", i.e., $<$5\% of their 1000 clones librate in the jovian 1:1 resonance for at least 1~kyr (red points in Figures~\ref{fig_ae} \& \ref{fig_ai_ei}). If an object had any clones that librate in the resonance for $\geq$1~kyr, the number of transient clones and their maximum resonant trap duration are provided. The 14 objects classified by the MPC/JPL as Trojans are shown in {\bf bold}.}\label{table:non-res}
\tablehead{\colhead{Designation} & \colhead{JPL Class} & \colhead{Transient Clones} & \colhead{Max Trap Duration (kyr)} }
\startdata
(944) A920 UB & Centaur & 0 & -- \\
(6144) 1994 EQ3\tablenotemark{1} & Asteroid & 0 & -- \\
{\bf (118624) 2000 HR24\tablenotemark{1}} & {\bf Trojan} & {\bf 0} & {\bf --} \\
(145485) 2005 UN398 & OMB & 0 & -- \\
(275618) 2000 AU242 & Asteroid & 0 & -- \\
(301964) 2000 EJ37\tablenotemark{1} & Asteroid & 0 & -- \\
(318875) 2005 TS100\tablenotemark{2} & Centaur & 0 & -- \\
(365756) 2010 WZ71 & Centaur & 4 & 29 \\
{\bf (371522) 2006 UG185\tablenotemark{1}} & {\bf Trojan} & {\bf 0} & {\bf --} \\
(380282) 2002 AO148 & Centaur & 0 & -- \\
(434762) 2006 HA153 & OMB & 0 & -- \\
(461363) 2000 GQ148 & Centaur & 0 & -- \\
(487496) 2014 SE288\tablenotemark{3} & Asteroid & 0 & -- \\
(490171) 2008 UD253 & Asteroid & 0 & -- \\
(494219) 2016 LN8 & Centaur & 0 & -- \\
(497619) 2006 QL39\tablenotemark{3} & Asteroid & 0 & -- \\
{\bf (497786) 2006 SA387\tablenotemark{3}} & {\bf Trojan} & {\bf 0} & {\bf --} \\
(498901) 2009 AU1 & OMB & 0 & -- \\
(504160) 2006 SV301 & Asteroid & 0 & -- \\
{\bf (515718) 2014 UQ194} &  {\bf Trojan} & {\bf 0} & {\bf --} \\
{\bf (517594) 2014 WX199} & {\bf Trojan} & {\bf 0} & {\bf --} \\
{\bf (528972) 2009 HM15} & {\bf Trojan} & {\bf 0} & {\bf --} \\
(529456) 2010 AN39 & OMB & 0 & -- \\
{\bf (576525) 2012 TQ67} & {\bf Trojan} & {\bf 0} & {\bf --} \\
(584530) 2017 GY10 & Asteroid & 0 & -- \\
(613349) 2006 BF208 & Centaur & 8 & 123 \\
(614590) 2009 XY21 & Asteroid & 0 & -- \\
{\bf 2000 CN152} & {\bf Trojan} & {\bf 0} & {\bf --} \\
2002 CF329 & Asteroid & 0 & -- \\
{\bf 2002 GE195\tablenotemark{3}} & {\bf Trojan} & {\bf 0} & {\bf --} \\
2005 NP82 & Centaur & 0 & -- \\
2005 TX214 & OMB & 0 & -- \\
2007 EV40\tablenotemark{3} & Asteroid & 0 & -- \\
2007 VW266  & Asteroid & 0 & -- \\
2009 SV412\tablenotemark{3} & Asteroid & 0 & -- \\
2010 BR88 & Asteroid & 0 & -- \\
{\bf 2010 BT86} & {\bf Trojan} & {\bf 6} & {\bf 1} \\
2010 CR140 & Centaur & 0 & -- \\
2010 ER22 & Asteroid & 0 & -- \\
2010 GP49 & Asteroid & 0 & -- \\
2010 JJ52 & Asteroid & 0 & -- \\
2010 KF52 & Asteroid & 0 & -- \\
2010 KG43 & Centaur & 0 & -- \\
2010 LV121 & Asteroid & 0 & -- \\
2010 RH69 & OMB & 0 & -- \\
2011 AF94 & Asteroid & 0 & -- \\
2011 AT15  & OMB & 0 & -- \\
2011 WM183 & Asteroid & 0 & -- \\
2012 DM127 & OMB & 0 & -- \\
2012 CM36 & Centaur & 0 & -- \\
2012 UJ38 & OMB & 0 & -- \\
2012 XO144 & Asteroid & 0 & -- \\
2013 AP182 & Asteroid & 0 & -- \\
2013 BU1 & Centaur & 0 & -- \\
2013 HA & OMB & 0 & -- \\
2013 KY14 & Asteroid & 0 & -- \\
2013 LA2 & Centaur & 0 & -- \\
2013 OL15 & Centaur & 5 & 6 \\
2013 VX9 & OMB & 1 & 2 \\
{\bf 2014 JK14} & {\bf Trojan} & {\bf 0} & {\bf --} \\
2014 JL128 & Asteroid & 2 & 1 \\
2014 KV3 & Centaur & 10 & 10 \\
2014 MA71 & Centaur & 0 & -- \\
2014 PA7 & Centaur & 0 & -- \\
2014 SZ398 & Asteroid & 0 & -- \\
2014 WY359 & Asteroid & 0 & -- \\
2015 AJ260 & Asteroid & 0 & -- \\
2015 BX306 & Centaur & 0 & -- \\
2015 CD60 & OMB & 0 & -- \\
2015 HO176 & Centaur & 0 & -- \\
2015 KM119 & OMB & 1 & 1 \\
2015 KW15 & Asteroid & 0 & -- \\
2015 MY90 & Centaur & 0 & -- \\
2015 OS110 & OMB & 0 & -- \\
2015 PC58 & Asteroid & 0 & -- \\
2015 PG119 & Centaur & 0 & -- \\
2015 VA53 & Asteroid & 0 & -- \\
2016 AH350 & Centaur & 0 & -- \\
2016 CE150\tablenotemark{3} & Asteroid & 0 & -- \\
2016 NG39 & Asteroid & 0 & -- \\
2016 PH135 & Asteroid & 0 & -- \\
2016 PW84 & Centaur & 0 & -- \\
2016 UV4 & Asteroid & 3 & 4 \\
2016 YB13 & Asteroid & 0 & -- \\
2017 DE104 & OMB & 0 & -- \\
2017 CD39 & Asteroid & 0 & - \\
2017 FP50 & Centaur & 0 & -- \\
2017 OY68 & Asteroid & 0 & -- \\
2017 TX13 & Asteroid & 0 & -- \\
2017 QO100\tablenotemark{3} & Centaur & 0 & -- \\
2017 WJ30\tablenotemark{3} & Asteroid & 0 & -- \\
2017 XJ65 & Asteroid & 0 & -- \\
2018 AN25 & Centaur & 0 & -- \\
2018 RG39 & Centaur & 0 & -- \\
2018 RH54 & OMB & 0 & -- \\
2018 VL10 & OMB & 0 & -- \\
2018 VX121 & OMB & 0 & -- \\
2018 XV35 & Centaur & 0 & -- \\
2019 KW30 & Asteroid & 0 & -- \\
2019 LM26 & Asteroid & 0 & -- \\
2019 PR2 & Amor & 0 & -- \\
2019 QS3 & Amor & 0 & -- \\
2019 QR6 & Amor & 0 & -- \\
2019 RF13 & Centaur & 0 & -- \\
{\bf 2019 SD164} & {\bf Trojan} & {\bf 33} & {\bf 3} \\
2019 SO48 & Centaur & 0 & -- \\
2020 BL76 & Centaur & 2 & 2 \\
{\bf 2020 BZ43} & {\bf Trojan} & {\bf 0} & {\bf --} \\
2020 HQ62 & Asteroid & 0 & -- \\
2020 PY28 & Asteroid & 0 & -- \\
{\bf 2020 YH25} & {\bf Trojan} & {\bf 0} & {\bf --} \\
2021 CD29 & Asteroid & 0 & -- \\
2021 CX19 & Asteroid & 0 & -- \\
2021 GT72 & Centaur & 32 & 15 \\
2021 JN58 & Centaur & 5 & 8 \\
2021 MJ1 & OMB & 0 & -- \\
2021 PM66\tablenotemark{3} & OMB & 0 & -- \\
2021 RZ47 & Centaur & 0 & -- \\
2021 VA28 & Centaur & 0 & -- \\
2021 UR & OMB & 5 & 26 \\
2021 WT6 & Centaur & 0 & -- \\
2022 BB25 & Centaur & 4 & 4 \\
2022 BF15 & Asteroid & 0 & -- \\
2022 CA31 & Centaur & 0 & -- \\
\textbf{Total} & \textbf{124} & & \\
Total Trojan & 14 & & \\
Total Asteroid & 48 & & \\
Total Centaur & 38 & & \\
Total OMB & 21 & & \\
Total Amor & 3 & & \\
\enddata
\tablenotetext{1}{Same classification as \cite{WajerKrolikowska2012}, \cite{DiRuzzaetal2023}, \cite{Karlsson2004}, and/or \cite{BeaugeRoig2001}.}
\tablenotetext{2}{\cite{WajerKrolikowska2012} classified (318875) 2005 TS100 as non-resonant, while \cite{DiRuzzaetal2023} classified it as a horseshoe.}
\tablenotetext{3}{\cite{WajerKrolikowska2012} and/or \cite{DiRuzzaetal2023} classify (487496) 2014 SE288, (497619) 2006 QL39, (497786) 2006 SA387, 2002 GE195, 2016 CE150, 2017 WJ30, and 2021 PM66 as horseshoes and \cite{DiRuzzaetal2023} classify 2007 EV40 as a quasi-satellite, 2009 SV412 as ``compound", and 2017 QO100 as ``transient", all based on their short-term orbital behavior over the next $\lesssim$1000 years. We require objects to remain stable in the 1:1 jovian resonance for $\geq$1~kyr in order to experience several libration periods before possible departure from the resonance (in the case of the transient captures); this is responsible for the classification differences for the objects that we classify as non-resonant that other studies find are resonant during the $\leq$1~kyr time scales they use.}
\end{deluxetable}




\clearpage

\end{document}